\documentclass[preprintnumbers,amsmath,amssymb]{revtex4-1}

\usepackage{epsfig}
\usepackage{graphicx}
\usepackage{subfigure}
\usepackage{float}
\usepackage{multirow}
\usepackage{amssymb}
\usepackage{amsmath}
\usepackage{picinpar}

\begin{document}

\newcommand{\figwidth}{0.8\columnwidth}
\newcommand{\subfigwidth}{0.6\columnwidth}
\newcommand{\subfigwidthtwo}{0.45\columnwidth}

\newcommand{\smin}{{\mbox{\scriptsize{min}}}}
\newcommand{\smax}{{\mbox{\scriptsize{max}}}}
\newcommand{\stot}{{\mbox{\scriptsize{tot}}}}
\newcommand{\tot}{{\mbox{\scriptsize{tot}}}}

\newcommand{\ms[2]}{{\mbox{\scriptsize{#2}}}}
\newcommand{\im}{i}
\newcommand{\jm}{j}

\newcommand{\mA}{\langle m_A\rangle}
\newcommand{\mB}{{\langle m_B\rangle}}
\newcommand{\pmB}{{\langle pm_B\rangle}}
\newcommand{\m}{{\langle m_\tot\rangle}}

\title{Ferrimagnetism and compensation temperature in spin-$1/2$ Ising trilayers}

\author{I. J. L. Diaz}\email{ianlopezdiaz@gmail.com}
\affiliation{Departamento de F\'{i}sica,
Universidade Federal de Santa Catarina,
88040-900, Florian\'{o}polis, SC, Brazil}
\author{N. S. Branco}\email{nsbranco@fisica.ufsc.br}
\affiliation{Departamento de F\'{i}sica,
Universidade Federal de Santa Catarina,
88040-900, Florian\'{o}polis, SC, Brazil}

\date{\today}

\begin{abstract}
The mean-field and effective-field approximations are applied in the
study of magnetic and thermodynamic properties of a spin-$1/2$ Ising system
containing three layers, each of which is composed exclusively of
one out of two possible types of atoms, \textbf{A} or \textbf{B}.
The \textbf{A-A} and \textbf{B-B} bonds are ferromagnetic
while the \textbf{A-B} bonds are antiferromagnetic.
The occurrence of a compensation phenomenon is verified and the compensation
and critical temperatures are obtained as functions of the Hamiltonian parameters.
We present phase diagrams dividing the parameter space in regions
where the compensation phenomenon is present or absent
and a detailed discussion about the influence of each parameter on the overall behavior of the system is made.
\end{abstract}


\maketitle

\section{Introduction}\label{introduction}

The interest in studies on ferrimagnets has increased considerably in the last few decades,
particularly due to a number of phenomena associated with these systems that present
great potential for technological applications \cite{connell1982magneto, camley1989theory, felser2007spintronics, phan2007review}.
Since the discovery  of ferrimagnetism in 1948 \cite{cullity2011introduction},
several theoretical models have been proposed to explain their magnetic behavior \cite{syozi1955statistical, hattori1966ising}.
Essentially, in these models the ferrimagnet is described as a combination of two or more
magnetically coupled substructures, e. g., sublattices, layers, or subsets of atoms within the system.
Each substructure may exhibit a different thermal behavior for its magnetization
and the combination of these different behaviors
may lead to the appearance of some interesting phenomena such as compensation points,
i. e., temperatures below the critical point for which the total magnetization is zero
while the individual substructures remain magnetically ordered \cite{cullity2011introduction}.

Mixed-spin Ising systems are often used as models to study ferrimagnetism.
The occurrence of compensation in such systems has been verified
in two-dimensional systems with a number of combinations of different spins (e. g. $s=1/2$, $1$, $3/2$, $2$, $5/2$)
\cite{godoy2000mixed, dakhama2000existence, nakamura2000existence, abubrig2001mean,
boechat2000ferrimagnetism, boechat2002renormalization, godoy2004mixed, ekiz2006effect, ekiz2006possibility, ekiz2006effect,
wang2015monte, wang2016effects, wang2016monte, wang2017compensation, wang2017monte, lv2017monte}.
Some single-spin systems,
such as layered magnets composed of stacked non-equivalent ferromagnetic planes,
have also been effectively used to model ferrimagnets.
A bilayer Ising system with spin-$1/2$ and no dilution has been studied via
transfer matrix (TM) \cite{lipowski1993layered, lipowski1998critical},
renormalization group (RG) \cite{hansen1993two, li2001critical, mirza2003phenomenological},
mean-field approximation (MFA) \cite{hansen1993two},
and Monte Carlo (MC) simulations \cite{ferrenberg1991monte, hansen1993two}.
Also the pair approximation (PA) has been applied to study
similar systems such as Ising-Heisenberg bilayers \cite{szalowski2012critical}
and multilayers \cite{szalowski2013influence} with spin-$1/2$ and no dilution.
Although site dilution is a crucial ingredient for the existence of a compensation point in a single-spin
system with even number of layers,
even with in the diluted systems the compensation effect will be present only under very specific conditions,
as has been verified through PA calculations for the Ising-Heisenberg bilayer \cite{balcerzak2014ferrimagnetism}
and multilayer \cite{szalowski2014normal}
and through MC simulations for the Ising bilayer \cite{diaz2017monte} and multilayer \cite{diaz2017multilayer}.

In contrast, for an odd number of layers, site dilution is no longer a necessary condition
for the existence of compensation in a single-spin system.
This was in fact confirmed in a very recent effective-field approximation (EFA) study
of Ising trilayer nanostructures \cite{santos2017effective}
for a few particular cases of Hamiltonian parameter values.
In order to study the conditions under which compensation effects may occur, we
propose a simple model, which is treatable by theoretical approximations and may
also be considerably easier to be built in experimental studies, when compared
to previous ferrimagnetic models.
The system we introduce in this work does not
require the presence of dilution, which is time- and memory-consuming in
numerical simulations and may be difficult to control in experimental set-ups.
More specifically, we study a three-layer Ising model with two types
of atoms (\textbf{A} and \textbf{B}, say), such that each layer is composed of only one type
of atom. Only two parameters are involved, the ratios between different
interactions. These parameters are changed, in order to establish the conditions
for the appearance of the compensation effect. We compare two different
theoretical approximations, which are easy to implement in the studied model.


The paper is organized as follows:
the theoretical model for the trilayer system is presented in Sec. \ref{sec:model},
in which we present the Hamiltonian for the system in Sec. \ref{sec:hamiltonian},
the mean-field analysis of the Hamiltonian in Sec. \ref{sec:MF},
and the effective-field analysis in Sec. \ref{sec:EF}.
The numerical results are presented and discussed in Sec. \ref{results}
and our conclusion and final remarks in Sec. \ref{conclusion}.

\section{Theoretical Model}\label{sec:model}

\subsection{Hamiltonian}\label{sec:hamiltonian}
\begin{figure}[h]
\begin{center}
\subfigure[\textbf{AAB}\label{fig:01:a}]{
\includegraphics[width=\subfigwidth]{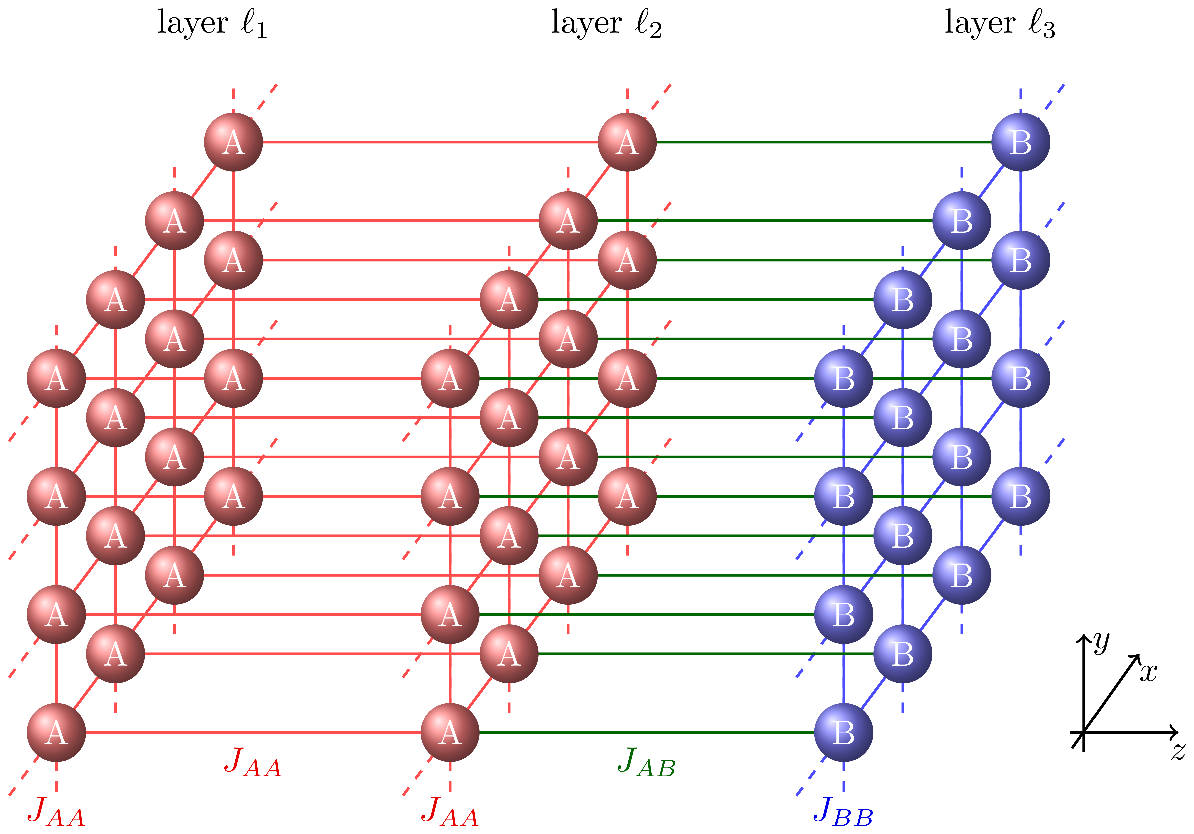}
}
\subfigure[\textbf{ABA}\label{fig:01:b}]{
\includegraphics[width=\subfigwidth]{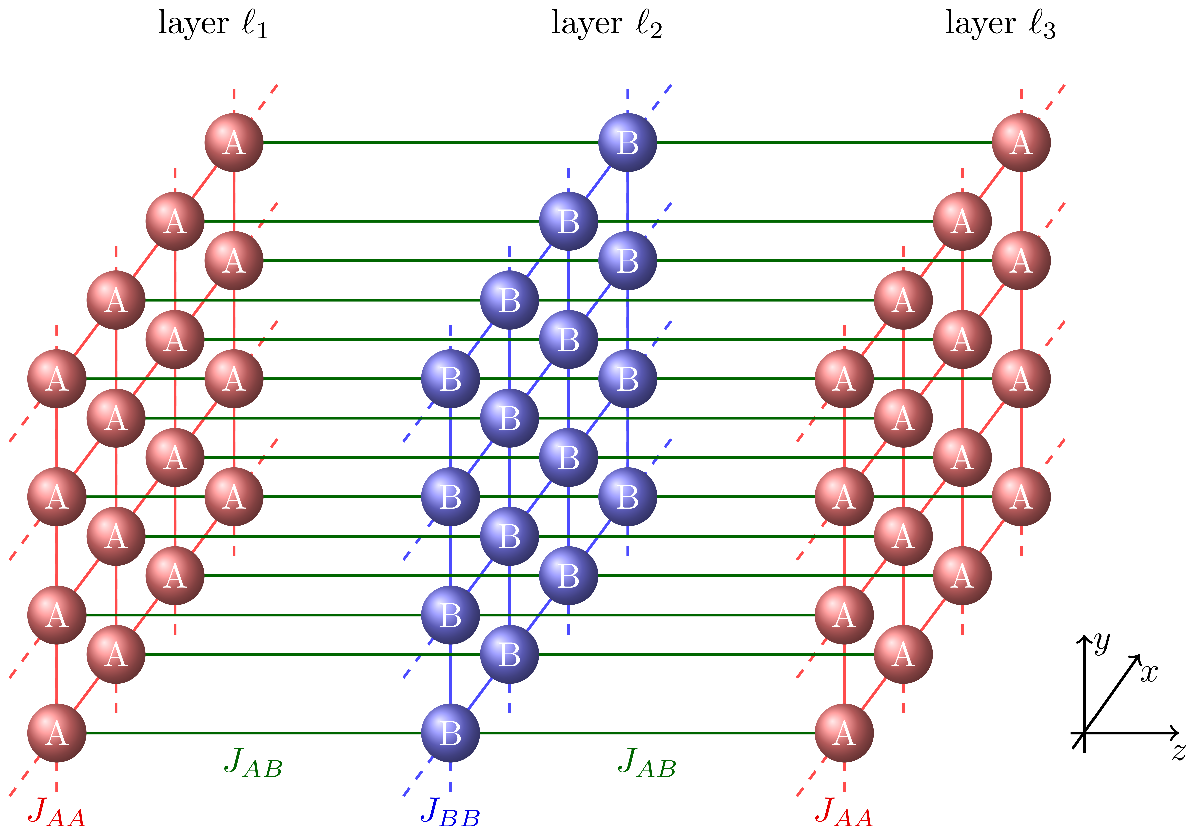}
}
\caption{
\label{fig:01}
A schematic representation of the trilayer systems.
In (a), we have the \textbf{AAB} system, in which
$J_{11}=J_{12}=J_{22}=J_{AA}>0$, $J_{23}=J_{AB}<0$, and $J_{33}=J_{BB}>0$.
In (b), we have the \textbf{ABA} system, in which $J_{11}=J_{33}=J_{AA}>0$; $J_{12}=J_{23}=J_{AB}<0$; $J_{22}=J_{BB}>0$.
}
\end{center}
\end{figure}

The trilayer system we study consists of three monoatomic layers, $\ell_1$, $\ell_2$, and $\ell_3$.
Each layer is composed exclusively of either type-\textbf{A} or type-\textbf{B} atoms (see Fig. \ref{fig:01}).
The general system is described by the spin-1/2 Ising Hamiltonian
\begin{align}\label{eq:hamiltonian}
-\beta\mathcal{H}=
 \sum_{\langle i\in\ell_1,j\in\ell_1\rangle}K_{11}s_i s_j
+\sum_{\langle i\in\ell_2,j\in\ell_2\rangle}K_{22}s_i s_j
+\sum_{\langle i\in\ell_3,j\in\ell_3\rangle}K_{33}s_i s_j
+\sum_{\langle i\in\ell_1,j\in\ell_2\rangle}K_{12}s_i s_j
+\sum_{\langle i\in\ell_2,j\in\ell_3\rangle}K_{23}s_i s_j,
\end{align}
where the sums run over nearest neighbors, $\beta\equiv (k_BT)^{-1}$, $T$ is the temperature, $k_B$ is the Boltzmann constant,
and the spin variables $s_i$ assume the values $\pm 1$.
The couplings are $K_{\gamma\eta}\equiv \beta J_{\gamma\eta}$,
where the exchange integrals $J_{\gamma\eta}$ are $J_{AA}>0$ for \textbf{A-A} bonds,
$J_{BB}>0$ for \textbf{B-B} bonds, and $J_{AB}<0$ for \textbf{A-B} bonds.

In this work we consider the two possible configurations of the trilayer with more atoms of type-\textbf{A} than type-\textbf{B} (see Fig. \ref{fig:01}).
The \textbf{AAB} system is the case in which
$J_{11}=J_{12}=J_{22}=J_{AA}$, $J_{23}=J_{AB}$, and $J_{33}=J_{BB}$ (Fig. \ref{fig:01:a}),
whereas the \textbf{ABA} system corresponds to
$J_{11}=J_{33}=J_{AA}$, $J_{12}=J_{23}=J_{AB}$, and $J_{22}=J_{BB}$ (Fig. \ref{fig:01:b}).
In both cases we wish to calculate the magnetization in each layer,
$m_\gamma\equiv \langle s_{i\in\ell_\gamma}\rangle$, $\gamma=1,2,3$, as well as the total magnetization
\begin{equation}
m_\tot=\frac{1}{3}(m_1+m_2+m_3).
\end{equation}

\subsection{Mean-field approximation (MFA)}\label{sec:MF}

For our analysis of the Hamiltonian \eqref{eq:hamiltonian},
we start by using the Callen identity \cite{callen1963note}
so the magnetizations can be written as
\begin{equation}\label{eq:mag}
m_\gamma = \langle\tanh{(\beta E_{i\in\ell_\gamma})}\rangle
\end{equation}
where $\langle\cdots\rangle$ denotes the canonical thermal average,
and
\begin{align}\label{eq:Ei}
\beta E_{i\in\ell_1} &= K_{11}\sum_\delta s_{(i+\delta)\in\ell_1}+K_{12}s_{i\in\ell_2} \nonumber \\
\beta E_{i\in\ell_2} &= K_{22}\sum_\delta s_{(i+\delta)\in\ell_2}+K_{12}s_{i\in\ell_1}+K_{23}s_{i\in\ell_3} \nonumber \\
\beta E_{i\in\ell_3} &= K_{33}\sum_\delta s_{(i+\delta)\in\ell_3}+K_{23}s_{i\in\ell_2},
\end{align}
where $\gamma=1,2$ or $3$.
The sums $\sum_\delta s_{(i+\delta)\in\ell_\gamma}$
are over the $z$ nearest neighbors of the $i$-th site in layer $\ell_\gamma$,
considering only the neighbors in the same layer.
For the particular case of square lattices, we have $z=4$.

In the standard mean-field approach, we have $\langle\tanh{(\beta E_{i\in\ell_\gamma})}\rangle = \tanh{\langle\beta E_{i\in\ell_\gamma}\rangle}$,
such that the means in Eq. \eqref{eq:mag} become
\begin{align}\label{eq_tgh}
m_1 &=
\tanh{(zK_{11}m_1 + K_{12}m_2)},
\nonumber \\
m_2 &=
\tanh{(zK_{22}m_2 + K_{12}m_1 + K_{23}m_3)},
\nonumber \\
m_3 &=
\tanh{(zK_{33}m_3 + K_{23}m_2)}.
\end{align}

The system \eqref{eq_tgh} was solved numerically to determine the magnetizations $m_1$, $m_2$, $m_3$, and $m_\tot$
as functions of the temperature for various values of the Hamiltonian parameters.
The results are presented in Sec. \ref{results}.

\subsection{Effective-field approximation (EFA)}\label{sec:EF}

In order to improve the mean-field approximation results
we employ the effective-field method first proposed by Honmura and Kaneyoshi \cite{honmura1979contribution}.
In this approach we use the differential operator
$e^{\lambda D}f(x)=f(x+\lambda)$, where $D\equiv\frac{\partial}{\partial x}$,
to we rewrite Eq. \eqref{eq:mag} as
\begin{equation}\label{eq:mag:EF}
m_\gamma = \langle \exp{(\beta E_{i\in\ell_\gamma}D)} \rangle \left.\tanh{x}\right|_{x=0},
\end{equation}
where $\gamma=1,2$ or $3$, and the $\beta E_{i\in\ell_\gamma}$ are given by Eqs. \eqref{eq:Ei}.

Substituting \eqref{eq:Ei} into Eq. \eqref{eq:mag:EF}, expanding the exponentials, and using the identities:
$(s_i)^{2n}=1$ and $(s_i)^{2n+1}=s_i$,
we obtain the following exact relations:
\begin{align}\label{eq:mags:EF:0}
m_1
=& \langle\Pi_\delta \{\cosh{(K_{11}D)}+s_{(i+\delta)\in\ell_1}\sinh{(K_{11}D)}\}
\times \{\cosh{(K_{12}D)}+s_{i\in\ell_2}\sinh{(K_{12}D)}\}\rangle \left.\tanh{(x)}\right|_{x=0},
\nonumber \\
m_2
=& \langle\Pi_\delta \{\cosh{(K_{22}D)}+s_{(i+\delta)\in\ell_2}\sinh{(K_{22}D)}\}
\times \{\cosh{(K_{12}D)}+s_{i\in\ell_1}\sinh{(K_{12}D)}\}
\nonumber \\&
\times \{\cosh{(K_{23}D)}+s_{i\in\ell_3}\sinh{(K_{23}D)}\}\rangle \left.\tanh{(x)}\right|_{x=0},
\nonumber \\
m_3
=& \langle\Pi_\delta \{\cosh{(K_{33}D)}+s_{(i+\delta)\in\ell_3}\sinh{(K_{33}D)}\}
\times \{\cosh{(K_{23}D)}+s_{i\in\ell_2}\sinh{(K_{23}D)}\}\rangle \left.\tanh{(x)}\right|_{x=0},
\end{align}
where again the $\delta$ index indicates the products are taken over the $z$ nearest neighbors of the $i$-th site.
After performing the thermal averages, neglecting multispin correlations
(i. e., $\langle s_is_j\cdots s_k\rangle= \langle s_i\rangle\langle s_j\rangle\cdots\langle s_k\rangle$),
and expanding the hyperbolic sines and cosines as exponentials,
it is possible to rewrite Eqs. \eqref{eq:mags:EF:0} as:
\begin{align}\label{eq:mags:EF:1}
m_1 =&
\frac{1}{2^{z+1}}\{(1+m_1)e^{K_{11}D}+(1-m_1)e^{-K_{11}D}\}^z 
\times \{(1+m_2)e^{K_{12}D}+(1-m_2)e^{-K_{12}D}\} \left.\tanh{(x)}\right|_{x=0} \nonumber \\
m_2 =&
\frac{1}{2^{z+2}}\{(1+m_2)e^{K_{22}D}+(1-m_2)e^{-K_{22}D}\}^z \nonumber \\
& \times \{(1+m_1)e^{K_{12}D}+(1-m_1)e^{-K_{12}D}\}
\times \{(1+m_3)e^{K_{23}D}+(1-m_3)e^{-K_{23}D}\}\left.\tanh{(x)}\right|_{x=0} \nonumber \\
m_3 =&
\frac{1}{2^{z+1}}\{(1+m_3)e^{K_{33}D}+(1-m_3)e^{-K_{33}D}\}^z 
\times \{(1+m_2)e^{K_{23}D}+(1-m_2)e^{-K_{23}D}\} \left.\tanh{(x)}\right|_{x=0}
\end{align}

As in Sec. \ref{sec:MF}, the system \eqref{eq:mags:EF:1} was solved numerically
to determine the magnetizations $m_1$, $m_2$, $m_3$, and $m_\tot$
as functions of the temperature for various values of the Hamiltonian parameters.
The results are presented in Sec. \ref{results}.

\section{Numerical Results and Discussion}\label{results}

\begin{figure}[h]
\begin{center}
\subfigure[With compensation (MFA).\label{fig:mags:AAB:MFA:a}]{
\includegraphics[width=\subfigwidthtwo]{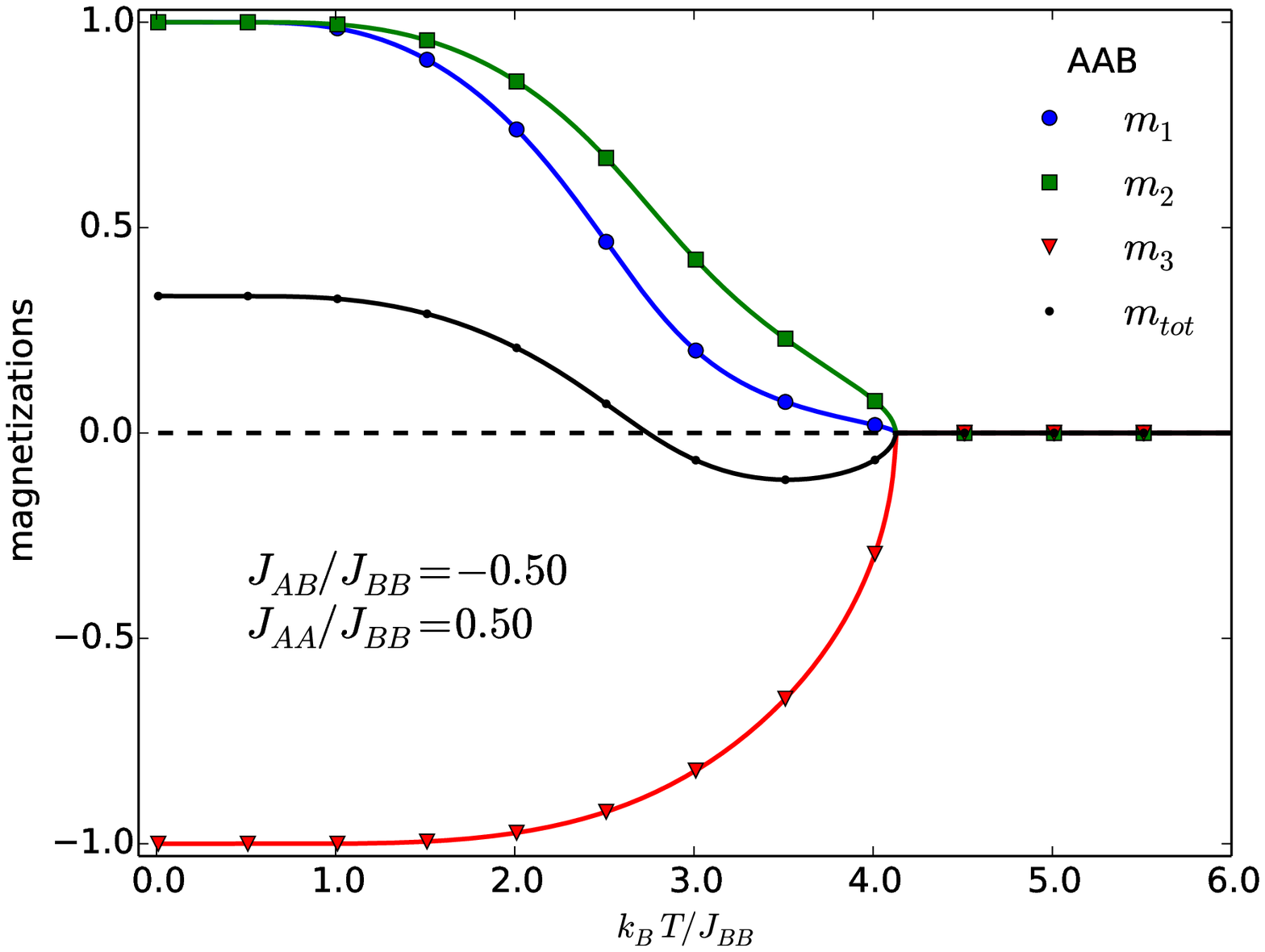}
}
\subfigure[Without compensation (MFA).\label{fig:mags:AAB:MFA:b}]{
\includegraphics[width=\subfigwidthtwo]{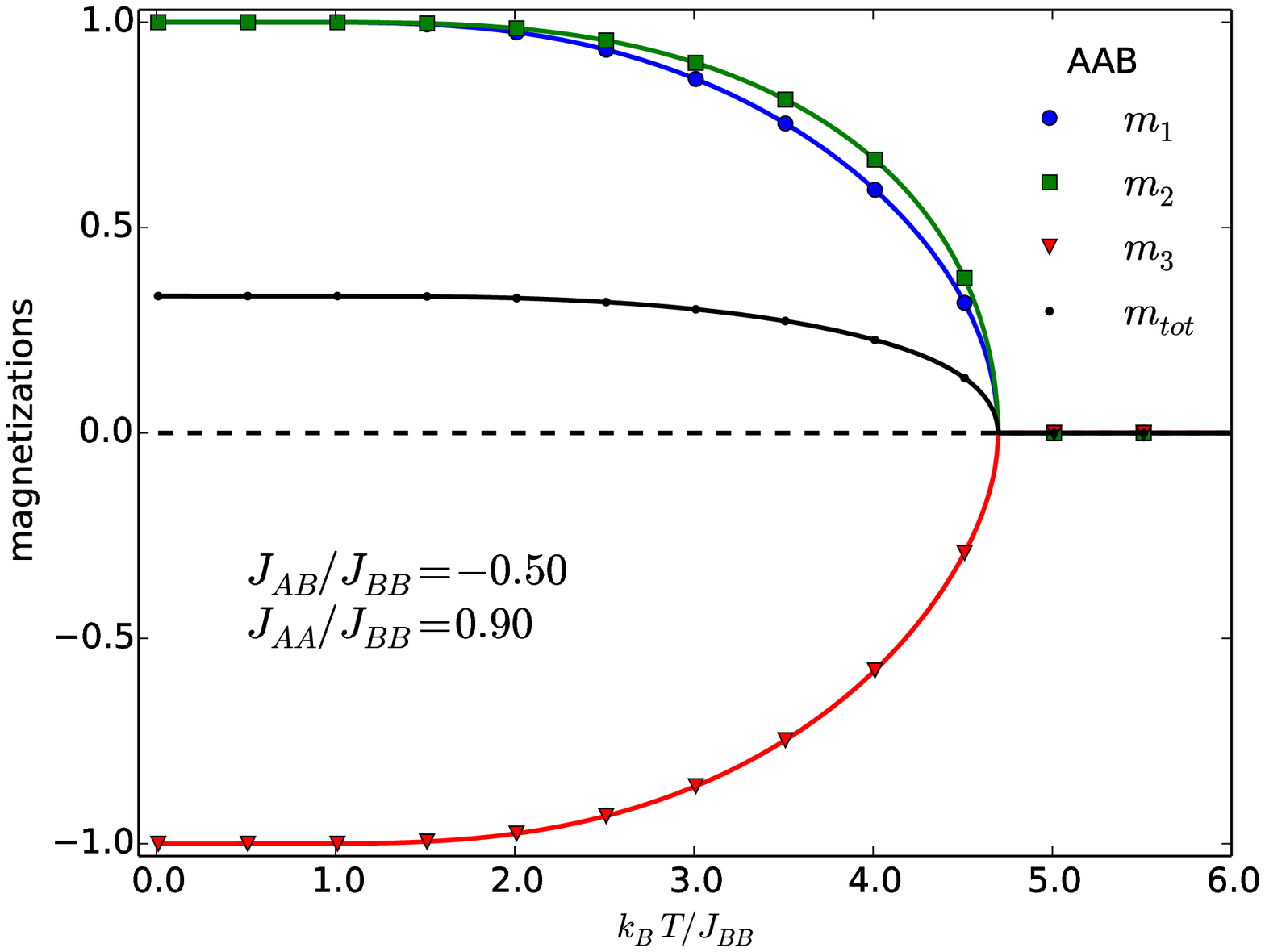}
}
\subfigure[With compensation (EFA).\label{fig:mags:AAB:EFA:a}]{
\includegraphics[width=\subfigwidthtwo]{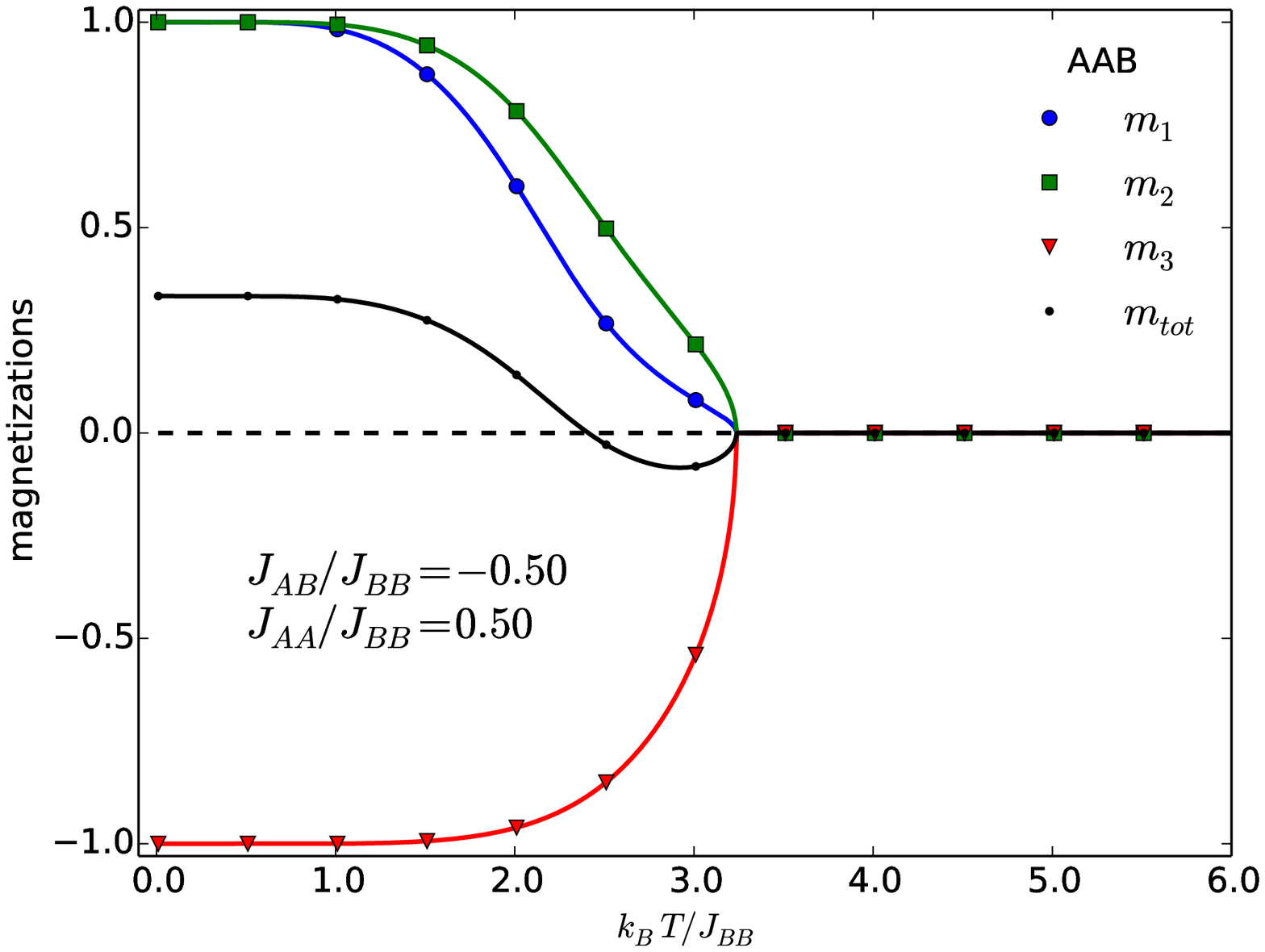}
}
\subfigure[Without compensation (EFA).\label{fig:mags:AAB:EFA:b}]{
\includegraphics[width=\subfigwidthtwo]{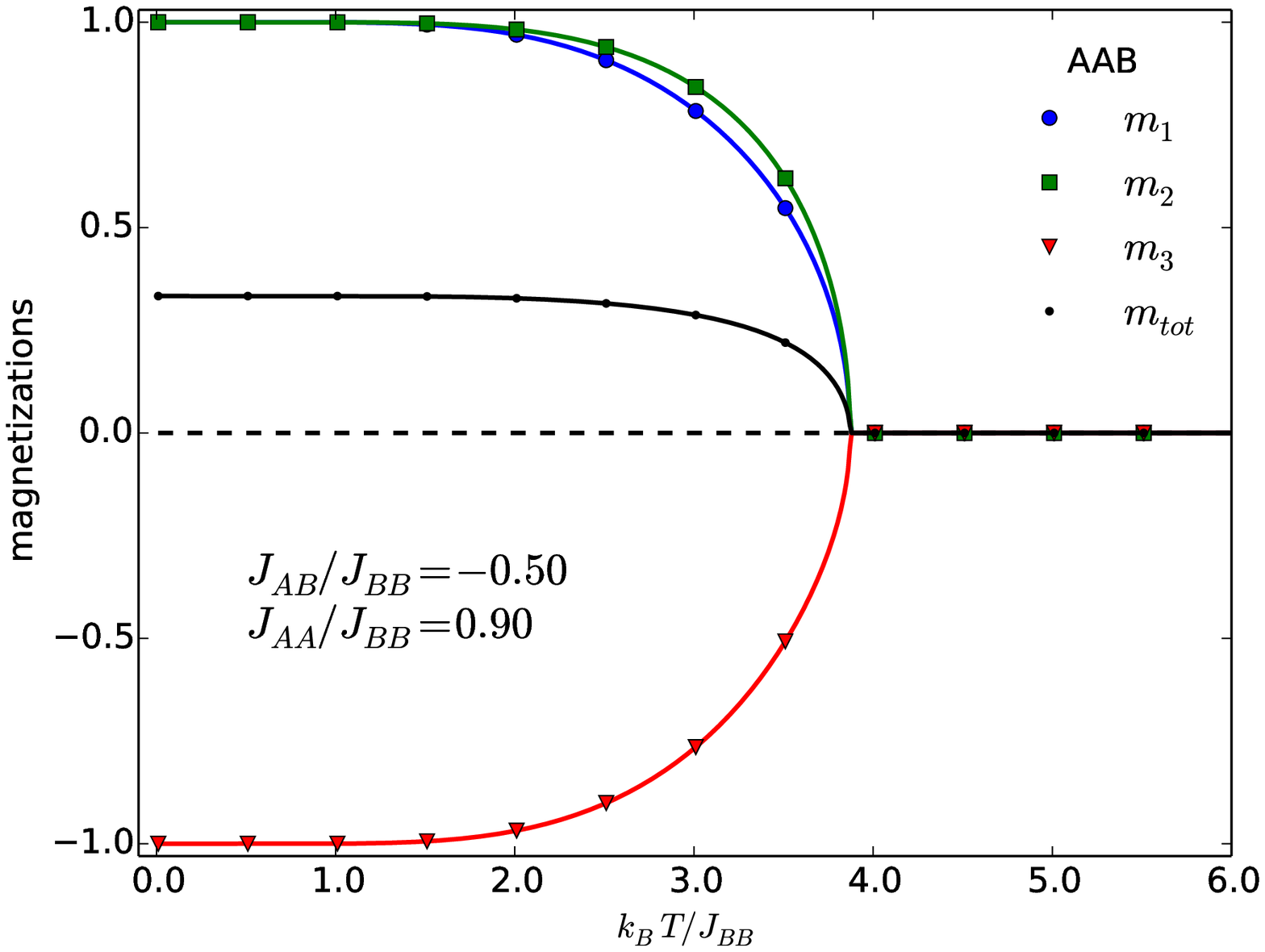}
}
\caption{
\label{fig:mags:AAB}
Magnetizations as functions of temperature for the \textbf{AAB} system
with $J_{AB}/J_{BB}=-0.50$.
For $J_{AA}/J_{BB}=0.50$,
both mean-field (a) and effective-field (c) approximations
show a compensation temperature $T_{comp}$ such that $m_{tot}=0$ and $0<T_{comp}<T_c$.
For $J_{AA}/J_{BB}=0.90$,
both mean-field (b) and effective-field (d) approximations
show no compensation effect.
}
\end{center}
\end{figure}

\begin{figure}[h]
\begin{center}
\subfigure[With compensation (MFA).\label{fig:mags:ABA:MFA:a}]{
\includegraphics[width=\subfigwidthtwo]{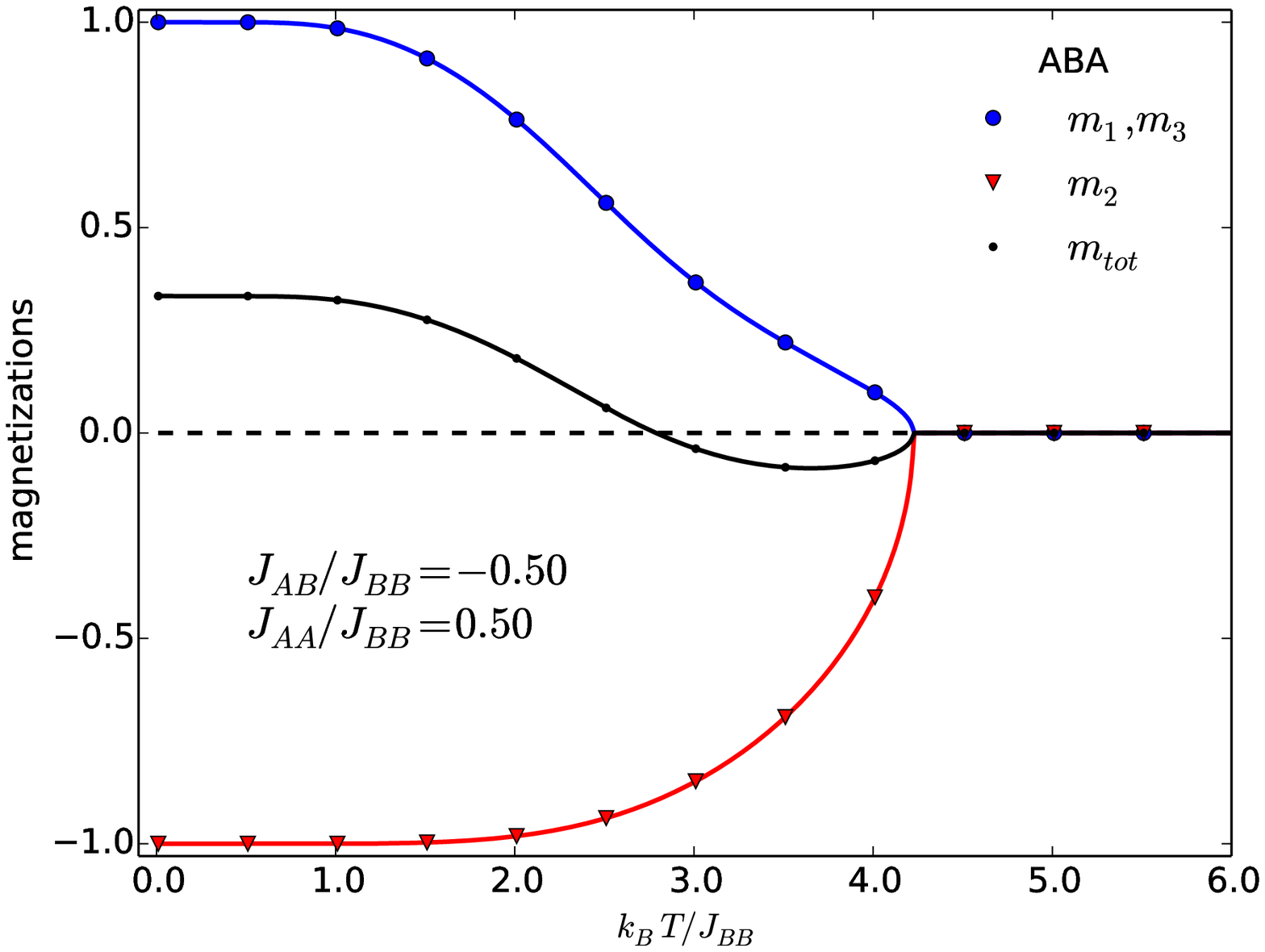}
}
\subfigure[Without compensation (MFA).\label{fig:mags:ABA:MFA:b}]{
\includegraphics[width=\subfigwidthtwo]{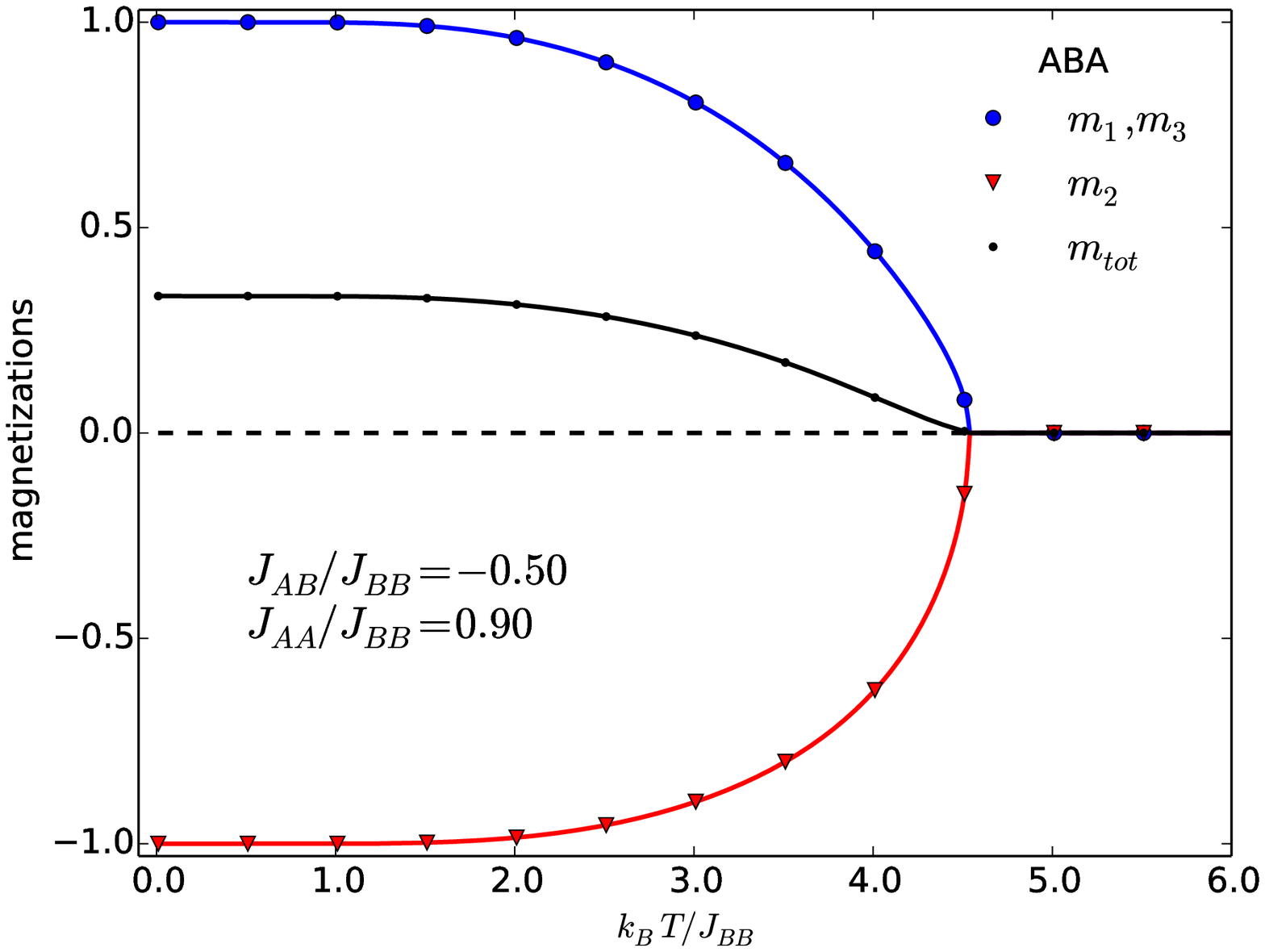}
}
\subfigure[With compensation (EFA).\label{fig:mags:ABA:EFA:a}]{
\includegraphics[width=\subfigwidthtwo]{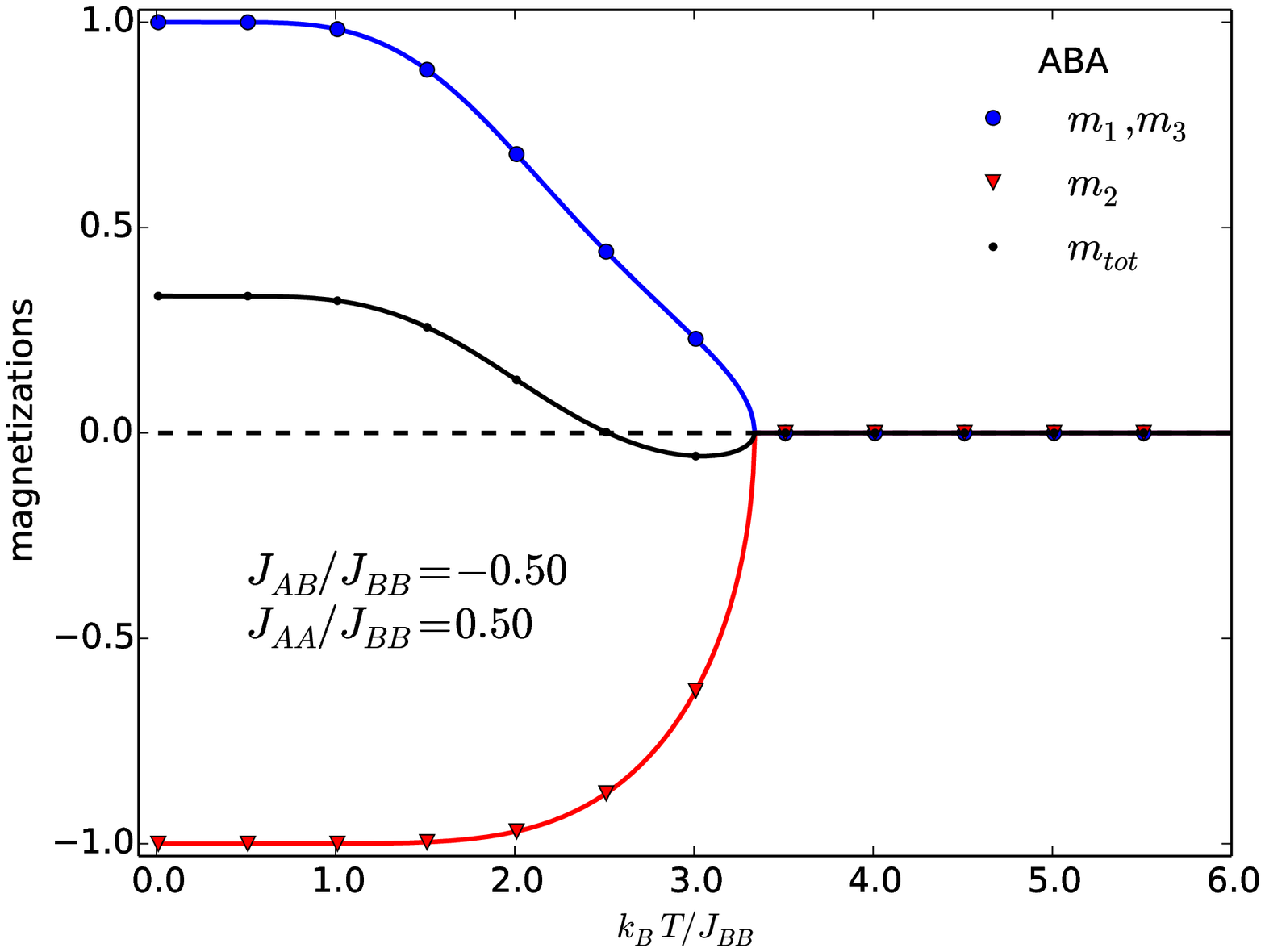}
}
\subfigure[Without compensation (EFA).\label{fig:mags:ABA:EFA:b}]{
\includegraphics[width=\subfigwidthtwo]{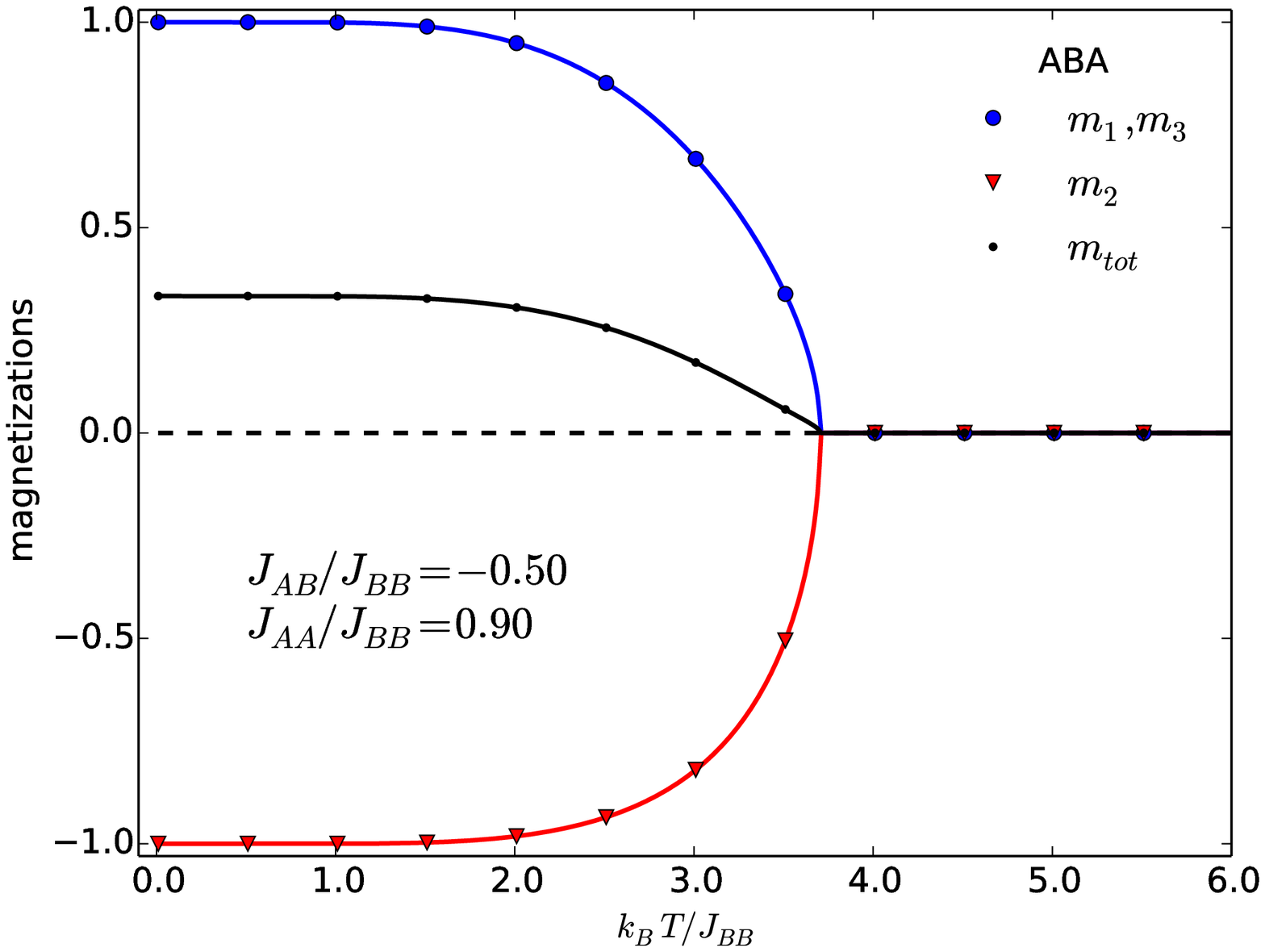}
}
\caption{
\label{fig:mags:ABA}
(Color online)
Magnetizations as functions of temperature for the \textbf{ABA} system
with $J_{AB}/J_{BB}=-0.50$.
For $J_{AA}/J_{BB}=0.50$,
both mean-field (a) and effective-field (c) approximations
show a compensation temperature $T_{comp}$ such that $m_{tot}=0$ and $0<T_{comp}<T_c$.
For $J_{AA}/J_{BB}=0.90$,
both mean-field (b) and effective-field (d) approximations
show no compensation effect.
}
\end{center}
\end{figure}

We start our analysis by solving the systems in Eqs. \eqref{eq_tgh} (MFA) and \eqref{eq:mags:EF:1} (EFA)
and looking at the temperature dependence of the magnetizations of the systems for
a range of values of the Hamiltonian parameters,
as shown in Figs. \ref{fig:mags:AAB} and \ref{fig:mags:ABA}.
The compensation point is determined for each set of Hamiltonian parameters as the temperature $T_{comp}$
for which $m_\tot=0$, while $m_1, m_2, m_3\neq 0$.
In turn, the critical point is determined as the temperature for which
all magnetizations vanish simultaneously.
Our goal in this work is to outline the contribution of each parameter
to the presence or absence of the compensation phenomenon.
To that end we map out the regions of the parameter space for which the system has
a compensation point, as seen in Figs. \ref{fig:mags:AAB:MFA:a} and \ref{fig:mags:ABA:MFA:a} for the MFA
and Figs. \ref{fig:mags:AAB:EFA:a} and \ref{fig:mags:ABA:EFA:a} for the EFA,
and the regions for which the compensation effect does not take place,
as seen in Figs. \ref{fig:mags:AAB:MFA:b} and \ref{fig:mags:ABA:MFA:b} for the MFA
and Figs. \ref{fig:mags:AAB:EFA:b} and \ref{fig:mags:ABA:EFA:b} for the EFA.

\begin{figure}[h]
\begin{center}
\subfigure[MFA.\label{fig:TvsJp:AAB:MFA}]{
\includegraphics[width=\subfigwidthtwo]{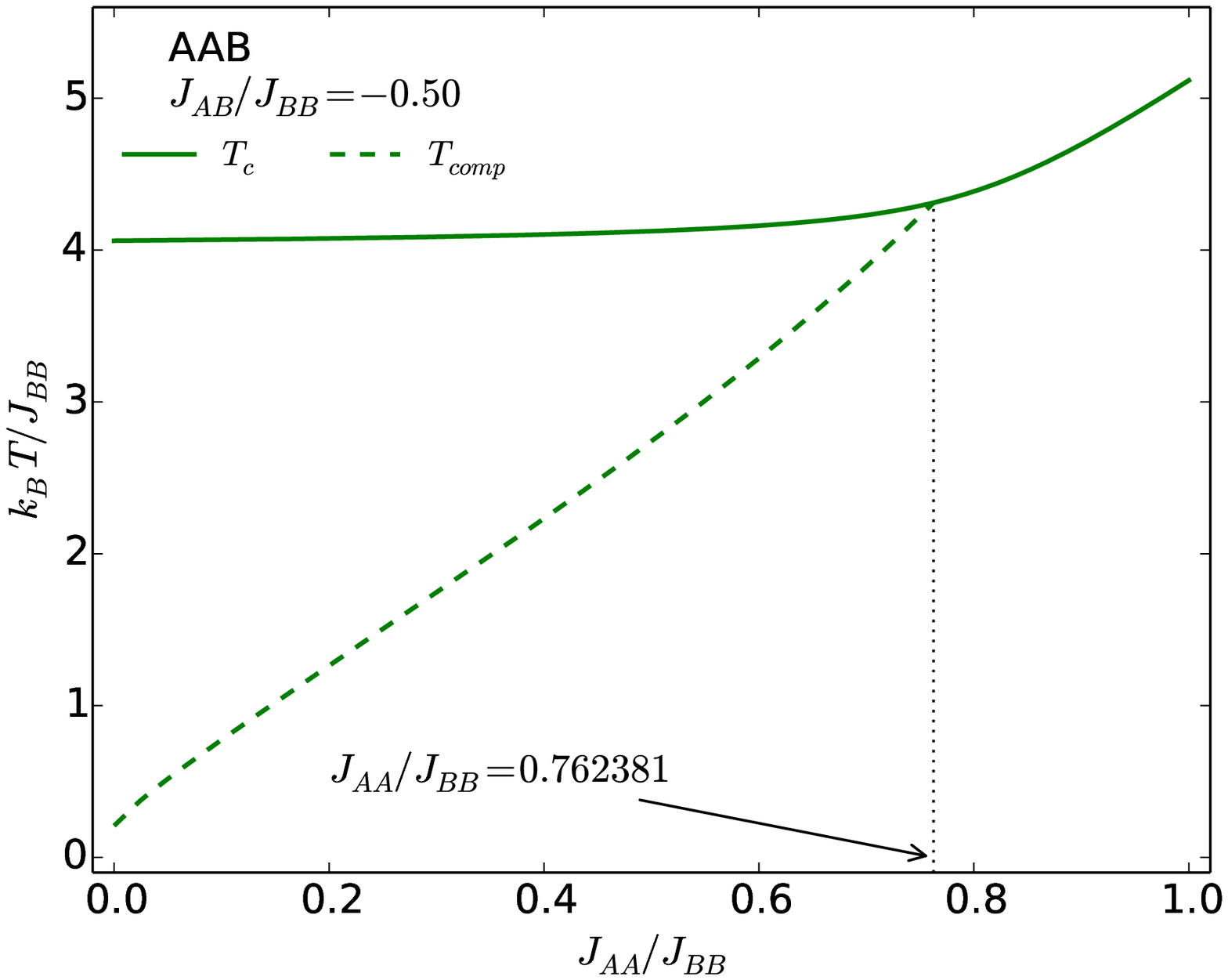}
}
\subfigure[EFA.\label{fig:TvsJp:AAB:EFA}]{
\includegraphics[width=\subfigwidthtwo]{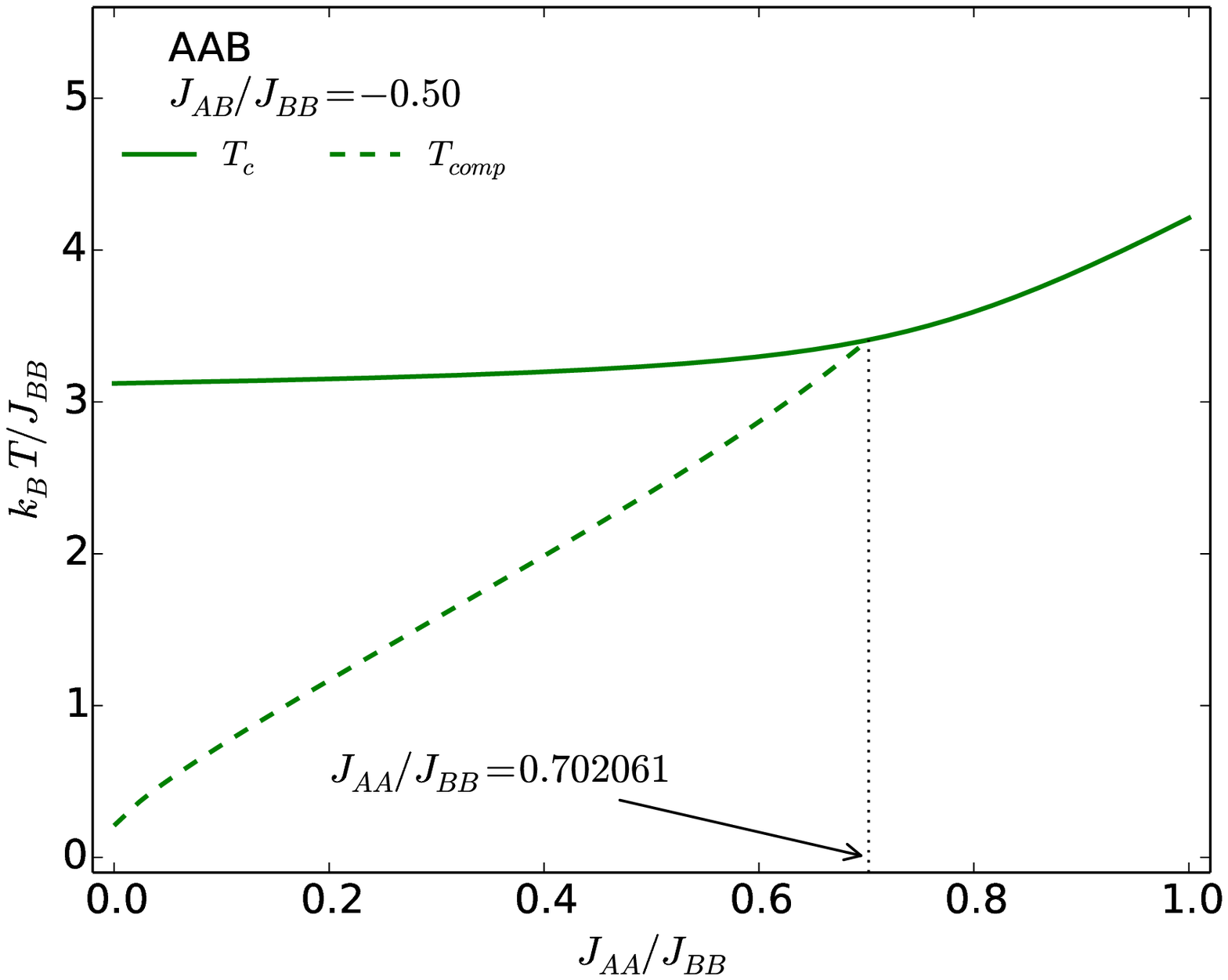}
}
\caption{
\label{fig:TvsJp:AAB}
$T_c$ (full line) and $T_{comp}$ (dashed line) as functions of $J_{AA}/J_{BB}$
for the \textbf{AAB} system with $J_{AB}/J_{BB}=-0.50$.
The dotted line marks the value of $J_{AA}/J_{BB}$ for which $T_{comp}=T_c$
and above which there is no compensation.
In (a) we present the results for the mean-field approximation, whereas
in (b) the results for the effective-field approximation are depicted.
}
\end{center}
\end{figure}

\begin{figure}[h]
\begin{center}
\subfigure[MFA.\label{fig:TvsJp:ABA:MFA}]{
\includegraphics[width=\subfigwidthtwo]{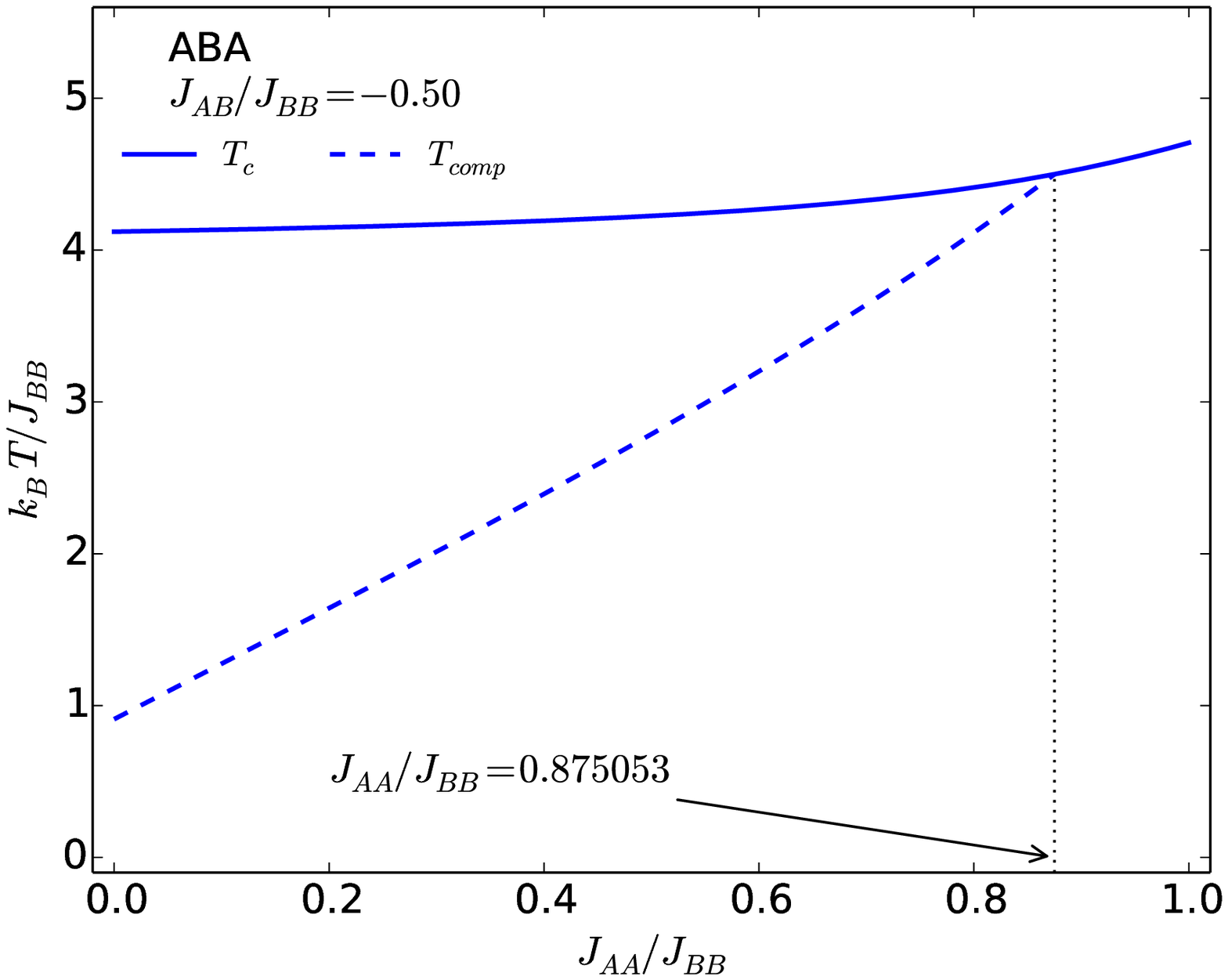}
}
\subfigure[EFA.\label{fig:TvsJp:ABA:EFA}]{
\includegraphics[width=\subfigwidthtwo]{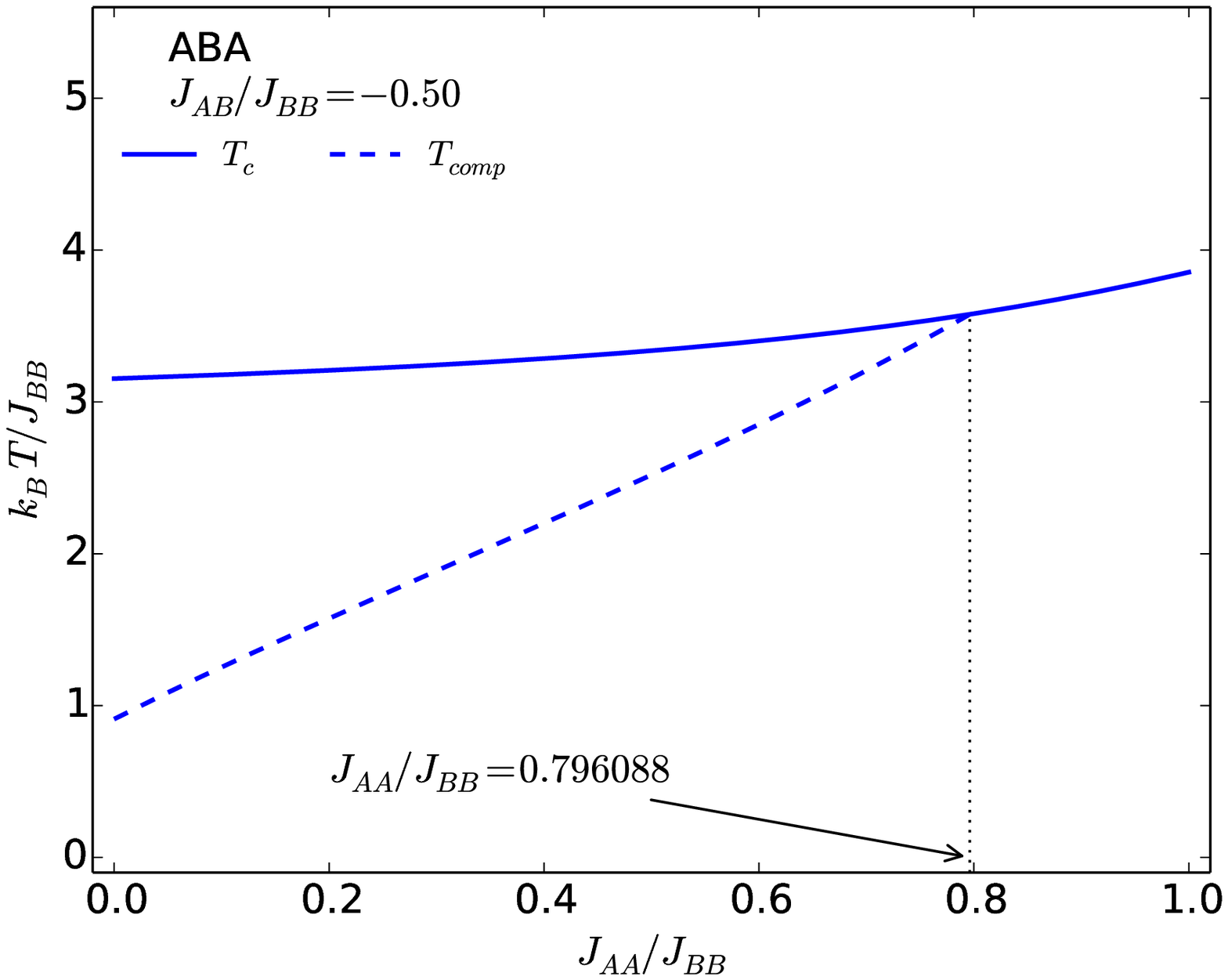}
}
\caption{
\label{fig:TvsJp:ABA}
$T_c$ (full line) and $T_{comp}$ (dashed line) as functions of $J_{AA}/J_{BB}$
for the \textbf{ABA} system with $J_{AB}/J_{BB}=-0.50$.
The dotted line marks the value of $J_{AA}/J_{BB}$ for which $T_{comp}=T_c$
and above which there is no compensation.
In (a) we present the results for the mean-field approximation, whereas
in (b) the results for the effective-field approximation are depicted.
}
\end{center}
\end{figure}

\begin{figure}[h]
\begin{center}
\subfigure[MFA.\label{fig:TvsJn:AAB:MFA}]{
\includegraphics[width=\subfigwidthtwo]{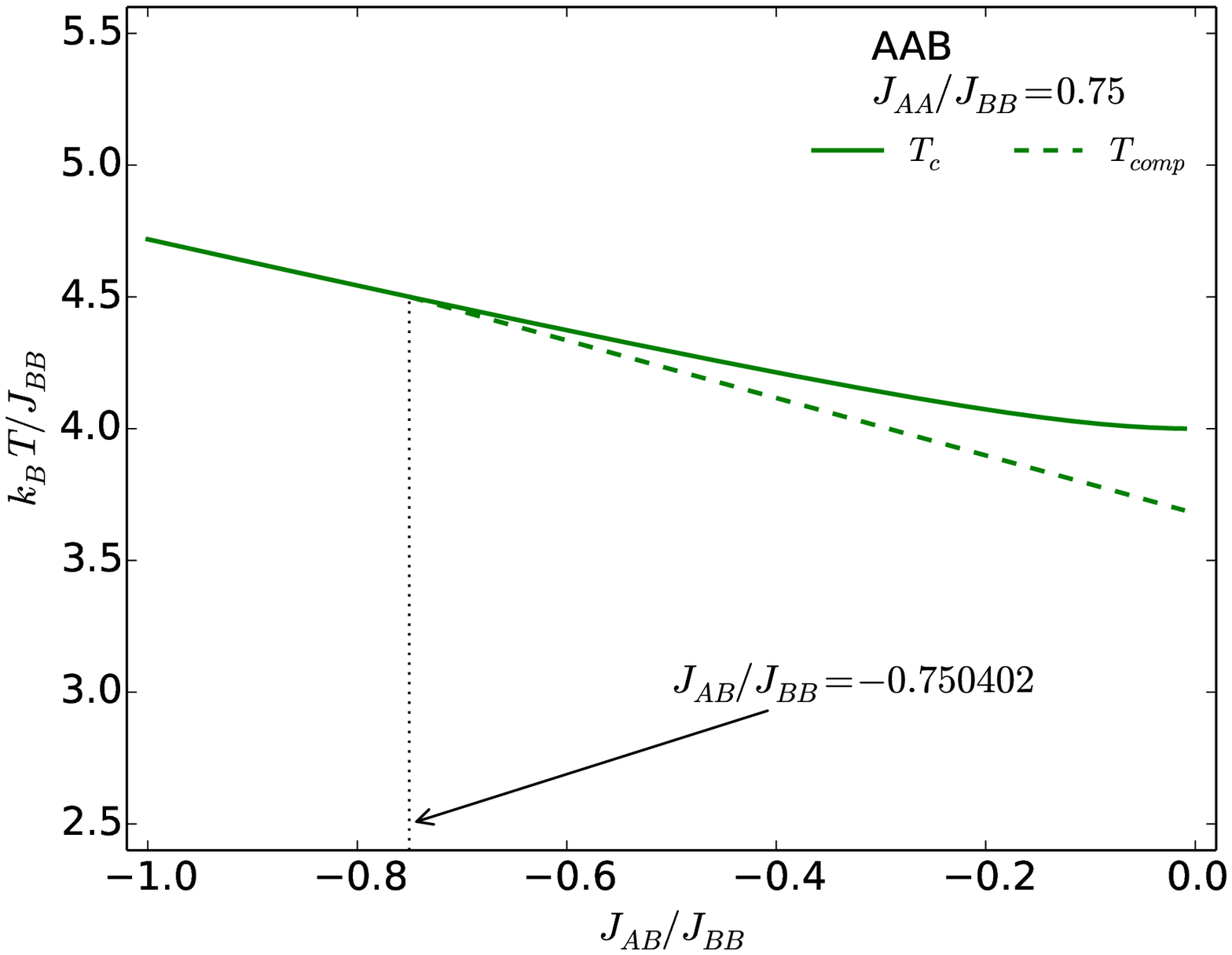}
}
\subfigure[EFA.\label{fig:TvsJn:AAB:EFA}]{
\includegraphics[width=\subfigwidthtwo]{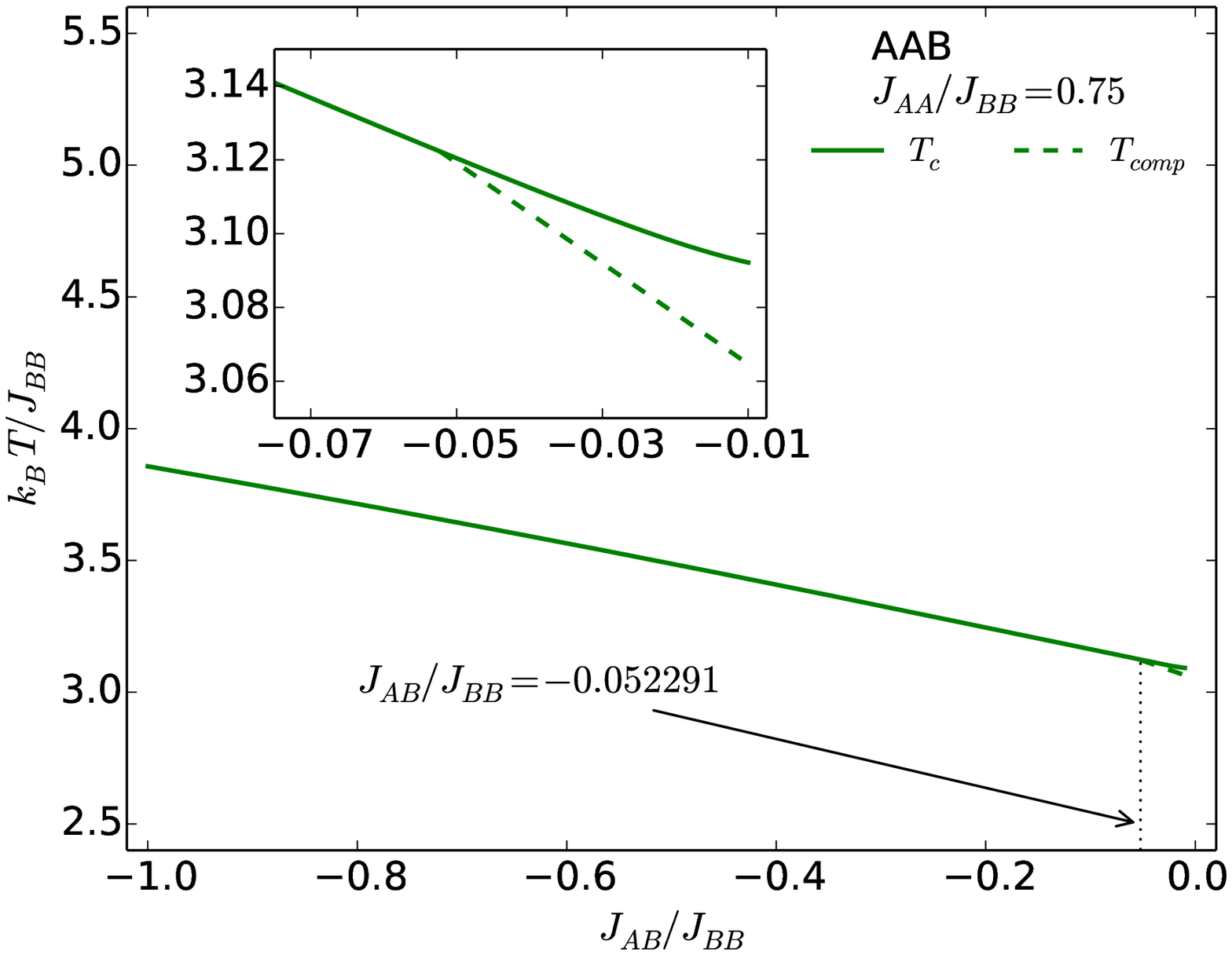}
}
\caption{
\label{fig:TvsJn:AAB}
$T_c$ (full line) and $T_{comp}$ (dashed line) as functions of $J_{AB}/J_{BB}$
for the \textbf{AAB} system with $J_{AA}/J_{BB}=0.75$.
The dotted line marks the value of $J_{AB}/J_{BB}$ for which $T_{comp}=T_c$
and below which there is no compensation.
In (a) we present the results for the mean-field approximation, whereas
in (b) the results for the effective-field approximation are depicted.
The inset in (b) is a zoom in the region where the $T_{comp}$ and $T_c$ curves meet.
}
\end{center}
\end{figure}

\begin{figure}[h]
\begin{center}
\subfigure[MFA.\label{fig:TvsJn:ABA:MFA}]{
\includegraphics[width=\subfigwidthtwo]{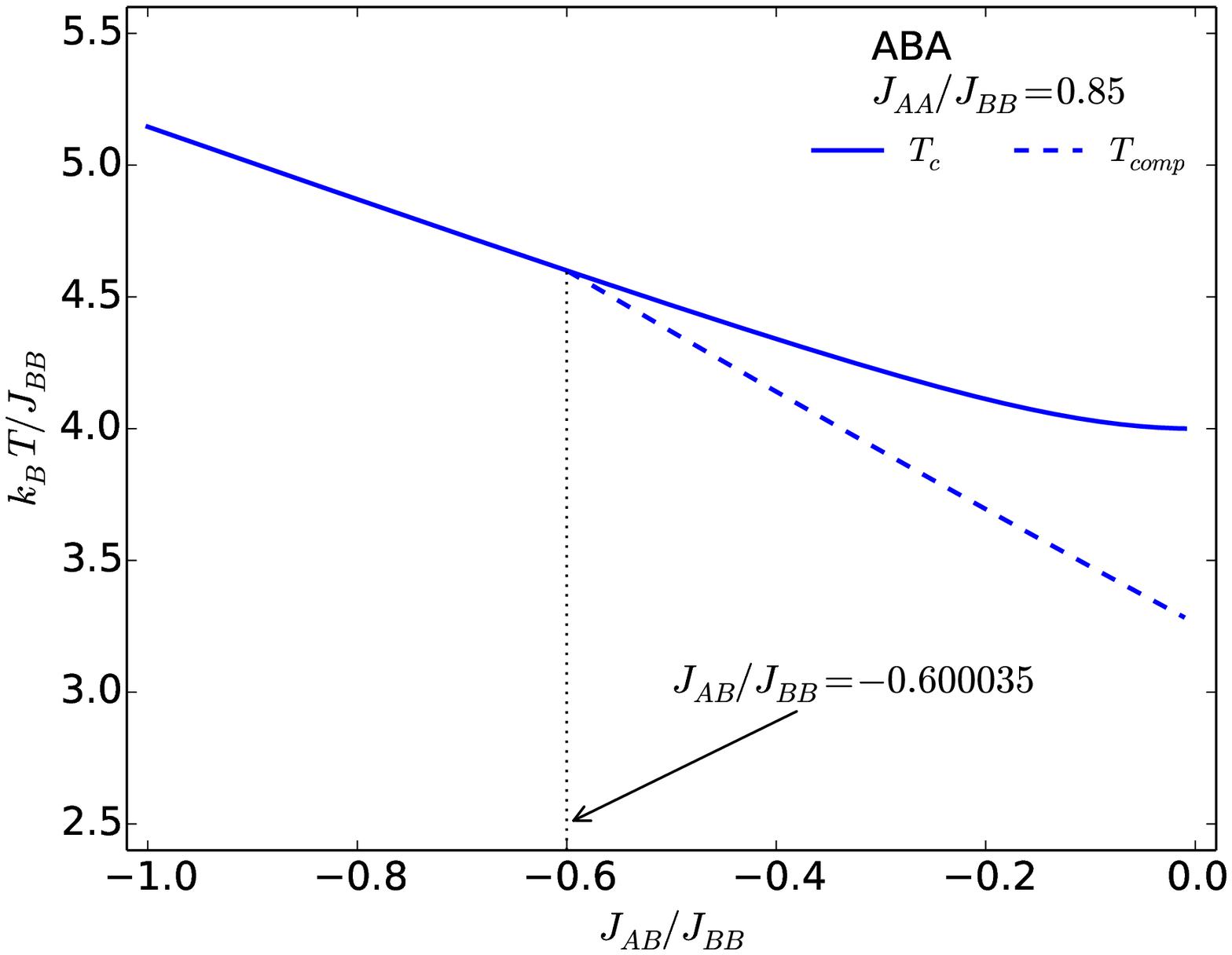}
}
\subfigure[EFA.\label{fig:TvsJn:ABA:EFA}]{
\includegraphics[width=\subfigwidthtwo]{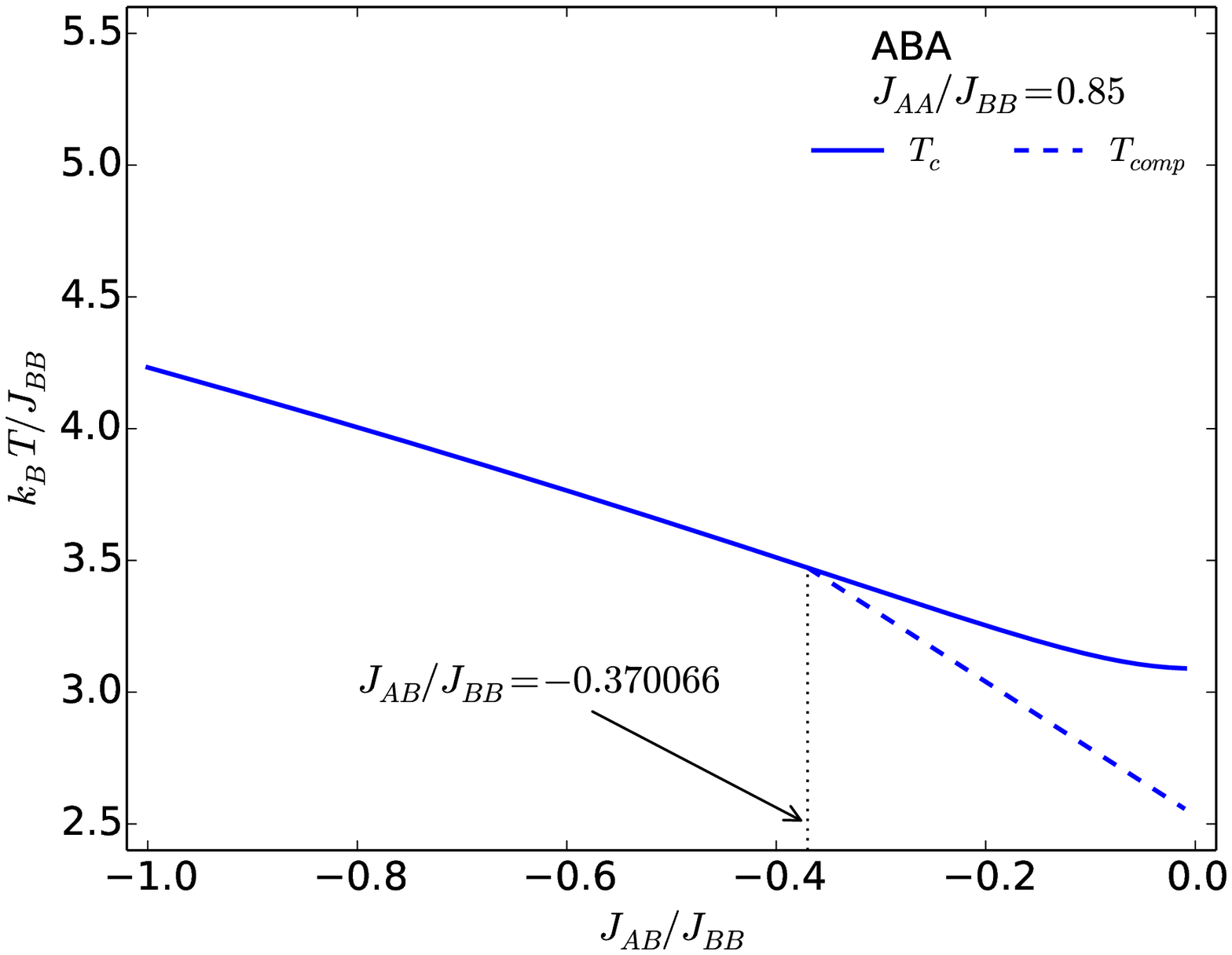}
}
\caption{
\label{fig:TvsJn:ABA}
$T_c$ (full line) and $T_{comp}$ (dashed line) as functions of $J_{AB}/J_{BB}$
for the \textbf{ABA} system with $J_{AA}/J_{BB}=0.85$.
The dotted line marks the value of $J_{AB}/J_{BB}$ for which $T_{comp}=T_c$
and below which there is no compensation.
In (a) we present the results for the mean-field approximation, whereas
in (b) the results for the effective-field approximation are depicted.
}
\end{center}
\end{figure}

In order to analyze the influence of $J_{AA}/J_{BB}$ in the behavior of the system,
we fix a value for $J_{AB}/J_{BB}$ and plot 
the critical temperatures and compensation temperatures as functions of $J_{AA}/J_{BB}$,
as seen
in Fig. \ref{fig:TvsJp:AAB} for the \textbf{AAB} system and
in Fig. \ref{fig:TvsJp:ABA} for the \textbf{ABA} system,
both cases for $J_{AB}/J_{BB}=-0.50$.
In Figs. \ref{fig:TvsJp:AAB:MFA} and \ref{fig:TvsJp:ABA:MFA} we have the results for the mean-field approximation, whereas
in Figs. \ref{fig:TvsJp:AAB:EFA} and \ref{fig:TvsJp:ABA:EFA} we have the results for the effective-field approximation.
In all cases, the dotted vertical lines mark the value of $J_{AA}/J_{BB}$ at which $T_c=T_{comp}$
and above which there is no compensation for each system.
Likewise, to understand the influence of $J_{AB}/J_{BB}$ in the behavior of the trilayers,
we fix a value for $J_{AA}/J_{BB}$ and obtain 
$T_c$ and $T_{comp}$ as functions of $J_{AB}/J_{BB}$,
as shown
in Fig. \ref{fig:TvsJn:AAB} for an \textbf{AAB} trilayer with $J_{AB}/J_{BB}=0.75$,
as well as
in Fig. \ref{fig:TvsJn:ABA} for an \textbf{ABA} trilayer with $J_{AB}/J_{BB}=0.85$.
In Figs. \ref{fig:TvsJn:AAB:MFA} and \ref{fig:TvsJn:ABA:MFA} we have the results for the mean-field approximation, whereas
in \ref{fig:TvsJn:AAB:EFA} and \ref{fig:TvsJn:ABA:EFA} we have the results for the effective-field approximation.
The dotted vertical lines mark the value of $J_{AB}/J_{BB}$ at which $T_c=T_{comp}$
and below which there is no compensation for each system.
The inset in Fig. \ref{fig:TvsJn:ABA:EFA} is a zoom in the region where the $T_{comp}$ and $T_c$ curves meet.

One important aspect about the comparison between the MFA and EFA results
is that the values of $T_c$ and $T_{comp}$ are consistently higher for the MFA than for the EFA.
For instance,
Figs. \ref{fig:mags:AAB} and \ref{fig:mags:ABA} show that
for $J_{AA}/J_{BB}=0.50$ and $J_{AB}/J_{BB}=-0.50$
the MFA critical temperature is $\approx 27\%$ higher than the EFA estimate
for both \textbf{AAB} and \textbf{ABA} systems.
When we increase $J_{AA}/J_{BB}$ to $0.90$ while keeping $J_{AB}/J_{BB}$ constant,
that percentile difference decreases to $\approx 21\%$ and $\approx 22\%$
for the \textbf{AAB} and \textbf{ABA} systems, respectively.
Although the difference is slightly less pronounced, for $J_{AA}/J_{BB}=0.50$ and $J_{AB}/J_{BB}=-0.50$,
the MFA compensation temperature is
$\approx 14\%$ ($\approx 11\%$) higher than the EFA estimate
for the \textbf{AAB} (\textbf{ABA}) trilayer.
This is expected since the effective-field theory takes into account short-range correlations,
which are entirely neglected by a standard mean-field approximation.
Therefore, although both methods overestimate the values of critical and compensation temperatures,
the values are expected to approach the true ones within the effective-field approximation framework.
It is worth stressing that the same occurs when we contrast
pair approximation \cite{balcerzak2014ferrimagnetism} and Monte Carlo \cite{diaz2017monte} results
for a site-diluted Ising bilayer,
in which case the PA temperatures are higher than the MC ones.
Although the PA takes into account longer-range correlations than both EFA and MFA,
it still systematically overestimates the temperatures since it is a mean-field-like approximation.
Monte Carlo simulations, on the other hand, do not neglect correlations and should therefore
provide temperature estimates that are much closer to the true values than their mean-field-like counterparts.

Similarly, by analyzing Figs. \ref{fig:TvsJp:AAB}, \ref{fig:TvsJp:ABA}, \ref{fig:TvsJn:AAB}, and \ref{fig:TvsJn:ABA},
we see that the percentile difference between the MFA and EFA estimates for the critical temperatures
is somewhere between $20\%$ and $35\%$,
being greater for both small $J_{AA}/J_{BB}$ and small $|J_{AB}/J_{BB}|$.
On the other hand, for the compensation temperatures
we have percentile differences between MFA and EFA estimates
ranging from $\approx 0\%$ to $30\%$, being greater for $J_{AA}/J_{BB}\rightarrow 1$
and for $|J_{AB}/J_{BB}|\rightarrow 0$.
Another important difference between mean-field and effective-filed results,
which also follows from Figs. \ref{fig:TvsJp:AAB}, \ref{fig:TvsJp:ABA}, \ref{fig:TvsJn:AAB}, and \ref{fig:TvsJn:ABA},
is that the area of these diagrams occupied by the ferrimagnetic phase with compensation
is smaller for the EFA than it is for the MFA.

\begin{figure}[h]
\begin{center}
\subfigure[MFA.\label{fig:phase:MFA}]{
\includegraphics[width=\figwidth]{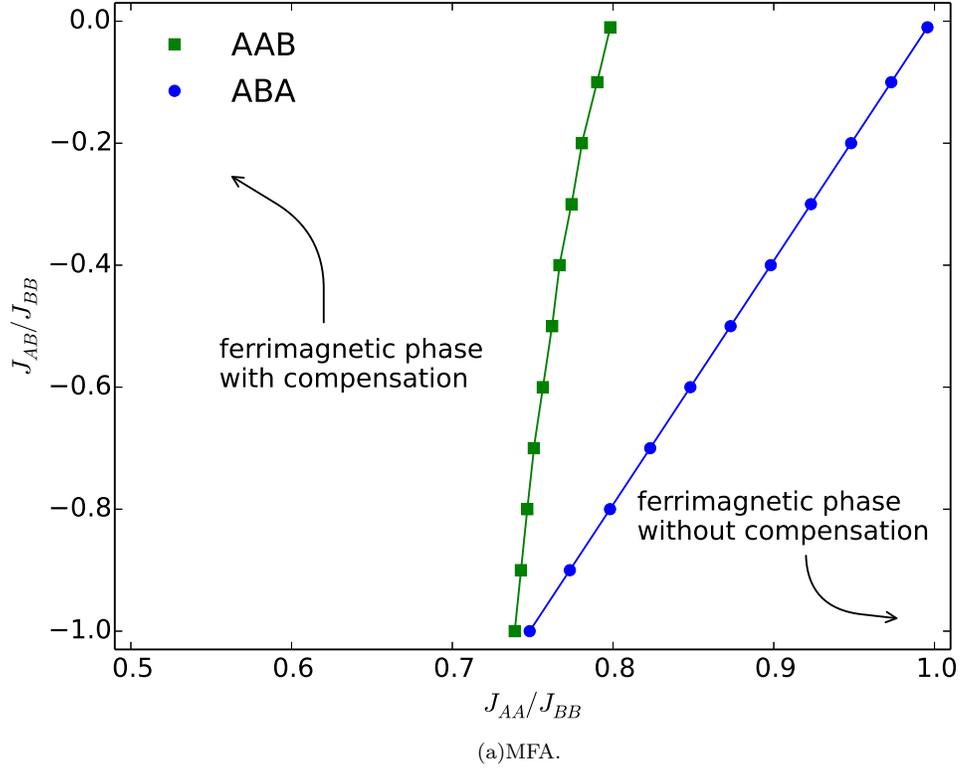}
}
\subfigure[EFA.\label{fig:phase:EFA}]{
\includegraphics[width=\figwidth]{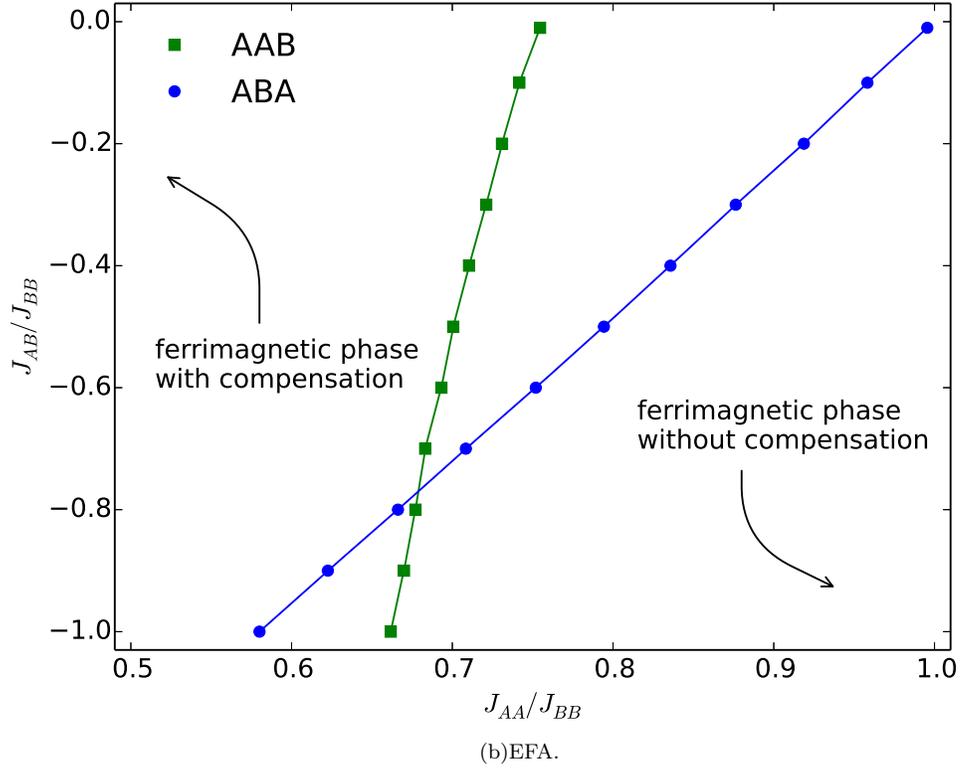}
}
\caption{
\label{fig:phase}
Phase diagrams for both \textbf{AAB} and \textbf{ABA} systems.
In (a) we present the results for the mean-field approximation, whereas
in (b) we present the results for the effective-field approximation.
}
\end{center}
\end{figure}

Finally, as it follows from the analyzes presented above,
it is convenient to divide the parameter space of our Hamiltonian in two distinct regions of interest.
One is a ferrimagnetic phase for which there is no compensation at any temperature
and the second is a ferrimagnetic phase where there is a compensation point at a certain temperature $T_{comp}$.
We present the results in Fig. \ref{fig:phase},
where we plot the phase diagrams for both \textbf{AAB} and \textbf{ABA} types of trilayer
and in both mean-field (Fig. \ref{fig:phase:MFA}) and effective-field (Fig. \ref{fig:phase:EFA}) approximations.
For each type of system, the line marks the separation between a ferrimagnetic phase with compensation (to the left)
and a ferrimagnetic phase without compensation (to the right).
These diagrams show that the compensation phenomenon will happen for a sufficiently small $J_{AA}/J_{BB}$
irrespective of the value of $J_{AB}/J_{BB}$,
although the range of values of $J_{AA}/J_{BB}$ for which the phenomenon occurs increases
as the \textbf{A-B} interplanar coupling gets weaker.
This behavior is similar to that of the diluted bilayer \cite{balcerzak2014ferrimagnetism, diaz2017monte}
and multilayer \cite{szalowski2014normal, diaz2017multilayer} systems for sufficiently small dilutions.

The main difference we see in Fig. \ref{fig:phase} between
systems \textbf{AAB} and \textbf{ABA} in both approximations
is that the \textbf{AAB} trilayer is less sensitive to the value of $J_{AB}/J_{BB}$ than the \textbf{ABA},
as the line separating the phases is more like a straight vertical line for the former system than for the latter.
This is consistent with the fact that the number of \textbf{A}-\textbf{B} bonds
in the \textbf{AAB} trilayer is only half that of the \textbf{ABA} trilayer.
In addition, Fig. \ref{fig:phase} shows that the area occupied by the ferrimagnetic phase with compensation in the $J_{AB}\times J_{AA}$ diagram
is smaller for the EFA than it is for the MFA for both types of trilayer,
confirming the trend seen in Figs. \ref{fig:TvsJp:AAB}, \ref{fig:TvsJp:ABA}, \ref{fig:TvsJn:AAB}, and \ref{fig:TvsJn:ABA}.
We see the same behavior when we compare the PA \cite{balcerzak2014ferrimagnetism} and MC \cite{diaz2017monte} results
for the Ising bilayer, in which case the smaller area is obtained through Monte Carlo simulations,
i. e., the area seems to decrease as we use more accurate approximations.
Thus, we expect that in future theoretical works on the trilayer systems,
the area of the phase with compensation will be smaller
in the PA and even smaller in MC simulations than what we obtained
in this work for both EFA and MFA.

\section{Conclusion}\label{conclusion}

We studied the thermodynamic and magnetic properties of an Ising trilayer model.
The system is composed of three planes, each of which can only have atoms of one out of two types (\textbf{A} or \textbf{B}).
The interactions between pairs of atoms of the same type (\textbf{A}-\textbf{A} or \textbf{B}-\textbf{B} bonds) are ferromagnetic
while the interactions between pairs of atoms of different types (\textbf{A}-\textbf{B} bonds) are antiferromagnetic.
The study is carried out through both a mean-field and an effective-field approaches.
The magnetic behavior of the system as a function of the temperature is obtained numerically.
We verified the occurrence of a compensation phenomenon and determined the compensation temperatures,
as well as the critical temperatures of the model for a range of values of the Hamiltonian parameters.

We present phase diagrams and a detailed discussion
about the conditions for the occurrence of the compensation phenomenon.
For instance, we see that the phenomenon is only possible if the $J_{AA}<J_{BB}$
and that the range of values of $J_{AA}/J_{BB}$ for which there is compensation
increases as $|J_{AB}/J_{BB}|$ gets smaller,
as it is also the case for similar systems containing a mixture of ferromagnetic and
antiferromagnetic bonds \cite{balcerzak2014ferrimagnetism, szalowski2014normal, diaz2017monte}.
The summary of the results is presented in a convenient way on $J_{AB}\times J_{AA}$
diagrams for both types of trilayer and for both mean-field and effective-field approximations.
These diagrams separate the Hamiltonian parameter-space
in two distinct regions: one corresponding to a ferrimagnetic phase where
the system has a compensation point and the other
is a ferrimagnetic phase without compensation.

It is clear from these diagrams that the area of the parameter space occupied by the
ferrimagnetic phase with compensation is smaller in the EFA than it is in the MFA
for both types of trilayer.
Thus, in a more sophisticated approach, this area could be even smaller.
However, we believe the compensation effect obtained here is robust and
cannot be just an artifact of the mean-field-like methods used in this work.
The fact that there are more atoms of type \textbf{A} than \textbf{B} in the system,
coupled with the fact that the antiferromagnetic interaction between atoms of
different types
favors the antiparallel alignment of the spins of atoms \textbf{A} and \textbf{B}, 
causes the system to exhibit a remanent magnetization for $T\rightarrow 0$.
Since the three layers are coupled with non-null exchange integrals,
all three magnetizations will go to zero at the same critical temperature;
therefore it is expected that a careful choice of the Hamiltonian parameters
may lead to situations where the individual magnetizations cancel each other out
below the critical point.
Nevertheless,
a confirmation of the occurrence of the phenomenon by more sophisticated theoretical methods,
such as pair approximation or Monte Carlo simulations, as well as the experimental
realization of a trilayer system with characteristics similar to the model presented
in this work would be of great value.


\begin{acknowledgments}
We are indebted to Prof. Dr. Lucas Nicolao for suggestions and helpful discussions.
This work has been partially supported by Brazilian Agency CNPq.
\end{acknowledgments}


%

\newpage

\end{document}